\documentclass[preprint]{rsl}

\begin{document}
\title{Evaluation of South African Candidate Sites for an Expanded Event Horizon Telescope}
\author{Senkhosi Simelane$^1$, Roger Deane$^{1,2}$, Athol Kemball$^{3,1}$, Roelf Botha$^4$, Roufurd Julie$^4$, Keitumetse Molamu$^4$, Adrian Tiplady$^4$, and Aletha de Witt$^5$
\\
\footnotesize $^1$\textit{Wits Centre for Astrophysics, School of Physics, University of the Witwatersrand, Braamfontein, Johannesburg, 2017, South Africa}\\
$^2$\textit{Department of Physics, University of Pretoria, Hatﬁeld, Pretoria, 0028, South Africa}\\
$^3$\textit{Department of Astronomy, University of Illinois Urbana-Champaign, Urbana, 61801, USA}\\
$^4$\textit{South African Radio Astronomy Observatory, River Park Liesbeek Parkway, Settlers Way, Cape Town, 7705, South Africa}\\
$^5$\textit{Department of Science, Technology and Innovation, 671 Meiring Naud\'{e} Road, Pretoria, 0184, South Africa}}

\newcommand{\bfig}{\begin{figure*}}
\newcommand{\efig}{\end{figure*}}
\newcommand{\bsfig}{\begin{subfigure}}
\newcommand{\esfig}{\end{subfigure}}
\newcommand{\be}{\begin{equation}}
\newcommand{\ee}{\end{equation}}
\newcommand{\bea}{\begin{eqnarray}}
\newcommand{\eea}{\end{eqnarray}}
\newcommand{\bi}{\begin{itemize}}
\newcommand{\ei}{\end{itemize}}
\newcommand{\bc}{\begin{center}}
\newcommand{\ec}{\end{center}}
\newcommand{\bt}{\begin{table}}
\newcommand{\et}{\end{table}}
\newcommand{\btab}{\begin{tabular}}
\newcommand{\etab}{\end{tabular}}
\newcommand{\nn}{\nonumber}
\newcommand{\nind}{\noindent}
\newcommand{\messier}{M87$^*$}
\newcommand{\SagA}{Sgr\,A$^*$}
\newcommand{\uv}{$(u, v)$}
\newcommand{\sa}{SA-EHT}
\newcommand{\amt}{AMT}
\newcommand{\ngehtsim}{\texttt{ngehtsim}}
\newcommand{\am}{\texttt{am}}
\newcommand{\Slat}{$\mathrm{S_{lat}}$}
\newcommand{\Slon}{$\mathrm{S_{lon}}$}
\newcommand{\Salt}{$\mathrm{S_{alt}}$}
\newcommand{\Sloc}{(\Slat, \Slon, \Salt)}
\newcommand{\Psurf}{$P_{\mathrm{surf}}$}
\newcommand{\ehtplus}{EHT$+$Africa}

\maketitle

%
%

\begin{abstract}
  Global expansion of the Event Horizon Telescope (EHT) will see the strategic addition of antennas at new geographical locations, transforming the sensitivity and imaging fidelity of the $\lambda \sim 1$\,mm EHT array. A possible South African EHT station would leverage a strong geographical advantage, local infrastructure, and radio astronomy expertise, and have strong synergies with the Africa Millimetre Telescope in Namibia. We assessed three South African candidate millimetre sites using climatological simulations and antenna sensitivity estimates, and found at least two promising sites. These sites are comparable to some existing EHT stations during the typical April EHT observing window and outperform them during most of the year, especially the southern hemisphere winter. The results suggest that a strategically placed South African EHT station will have a sizable, positive impact on next-generation EHT objectives and the resulting black hole imaging science.
\end{abstract}

\section{Introduction}
A host of novel black hole (BH) science goals will be enabled by the expansion of the Event Horizon Telescope (EHT) \cite{Johnson2023, Doeleman2023}, including the reconstruction of high-fidelity movies of the Messier 87$^*$ (\messier) and Sagittarius A$^*$ (\SagA) supermassive BHs (SMBHs). Both sources display dynamic behaviour, with timescales of $\sim20$\,days \cite{Bower2015} and $\sim20$\,minutes \cite{Genzel2003, Ghez2004, Gravity2020}, respectively. However, the typical one-week duration of EHT observing campaigns is too short to capture the full dynamical evolution of \messier~\cite{EHT2019_v} and limited snapshot $(u, v)$-coverage has meant that the rapid time-variability of \SagA~remains poorly constrained \cite{EHT2022_i}. These limitations have precluded experiments that allow dynamical imaging of SMBHs. New and strategically placed stations, especially several on the African continent, would extend the annual monitoring window for \messier~and also significantly improve the instantaneous and full-track $(u, v)$-coverage towards \SagA~\cite{Deane2022, LaBella2023}, while prompting significant strides in African Very Long Baseline Interferometry (VLBI) development, particularly with intra-African VLBI baselines.\par

Southern Africa holds a strong geographical advantage for imaging southern-hemisphere sources, such as \SagA. Simulations by \cite{LaBella2023} showed that adding the Africa Millimetre Telescope (\amt) in Namibia and a Canary Islands station to the EHT improves \SagA~movie fidelity through better constraints on source extent and a factor $\gtrsim4$ increase in observation length with adequate snapshot \uv-coverage for dynamical imaging. Static \SagA~imaging from \cite{Deane2022} showed that an EHT array augmented with the \amt~and a South African station achieves significantly improved \uv-coverage, especially in the eastern sub-array. Through long north-south and east-west baselines to NOEMA and ALMA, respectively, a South African EHT station (\sa~hereafter) would further enhance dynamical reconstructions of \SagA~and \messier and potentially enable polarimetric photon ring experiments, as proposed by \cite{Palumbo2023}. Additionally, a short ($\sim1000$\,km) \sa-\amt~baseline would enable the measurement of low-spatial-frequency emission critical for linking jet launch with horizon-scale dynamics. The added short-baseline constraints would also improve the gain calibration solutions and hence the achieved image fidelity \cite{Cornwell1999} and dynamic range, particularly with Maximum Entropy Method imaging \cite{Narayan1986}.\par

The prospective array performance improvements offered by the addition of \sa~to the EHT strongly motivate an investigation of feasible millimetre (mm) sites in South Africa. To this end, three sites were identified for preliminary evaluation: Ben Macdhui (BMAC) in South Africa's Eastern Cape, Sutherland (STL) in the Northern Cape, and Matjiesfontein (MATJ) in the Western Cape. Details of the sites, selected for their relatively high altitude and existing infrastructure, are summarised in Table~\ref{tab:sa-eht_site_info}. The pre-selection criterion for sites to have existing infrastructure is motivated by a more rapid and cost-effective \sa~project delivery.\par
\begin{table}[h]
    \centering
    \small
    \setlength{\tabcolsep}{4pt}  
    \btab{|l|c|c|c|}
    \hline
        \textbf{Site} & \textbf{Latitude ($^\circ$)} & \textbf{Latitude ($^\circ$)} & \textbf{Altitude (m)} \\
        \hline
        Ben Macdhui & -30.648 & 27.935 & 3000  \\
        Sutherland & -32.378 & 20.813 & 1800 \\
       Matjiesfontein & -33.266 & 20.582 & 1340 \\
        \hline
    \etab
    \caption{\small Geographical information of the three \sa~candidate sites: Ben Macdhui (BMAC), Sutherland (STL) and Matjiesfontein (MATJ).}
    \label{tab:sa-eht_site_info}
\end{table}

Tropospheric water vapour is the chief cause of atmospheric absorption and scattering of mm waves, necessitating high-altitude, low-humidity sites. The frequency-dependent absorption, quantified by the optical depth, $\tau(\nu)$, attenuates incoming mm radiation as $e^{-\tau(\nu)}$ and a fast-evolving water vapour distribution above a mm station causes rapid phase fluctuations. The coherence time, defined as the mean time interval over which the troposphere induces a 1-radian phase shift, characterises the rate of these fluctuations. Thus, the column-integrated precipitable water vapour (PWV) and $\tau(\nu)$ are key parameters in mm site evaluations. For the next-generation EHT (ngEHT), the $86\,\mathrm{GHz}$, $230\,\mathrm{GHz}$, and $345\,\mathrm{GHz}$ zenith optical depths are of the most interest as these align with the proposed receiver bands \cite{Doeleman2023}. Wind speeds are also critical, as high wind speeds degrade coherence times, disrupt dish pointing and tracking and may even necessitate stowing of the dish. This work evaluated these climatological proxies, among others, and used them to estimate antenna sensitivity for various dish sizes via the System Equivalent Flux Density (SEFD). This analysis marks a first step towards identifying an SA-EHT dish size that offers the best value proposition while delivering significant scientific impact.\par

The work is presented as follows: the methods employed in the climatological site characterisation and antenna sensitivity estimation study are presented in Sections~\ref{sect:climatological_site_characterisation} and \ref{sect:sensitivity_estimation}. The results are presented and discussed in Section~\ref{sect:results_and_discussion}. Finally, Section~\ref{sect:conclusion} summarises the key results and gives an outlook on the next steps and outstanding questions.

\section{Climatological Modelling}\label{sect:climatological_site_characterisation}
To characterise the candidate \sa~sites, we employed a meteorological modelling approach akin to \cite{Roelofs2020, Raymond2021, Yu2023, Pesce2024}. Using data from Modern-Era Retrospective Analysis for Research and Applications, version 2 (MERRA-2) \cite{Gelaro2017}, a global database of atmospheric measurements on a $0.5\,^\circ$ latitude by $0.625\,^\circ$ longitude (or $\sim55$\,km latitude by $\sim70$\,km) grid from 1980 to date, we generated realistic model atmospheres. MERRA-2 has a three-hour temporal resolution, but a temporal resolution of $1$\,month was used for this analysis as it suffices for the goal of shortlisting sites for onsite testing. We used daily MERRA-2 data files from 01-01-2009 to 31-12-2022 to calculate monthly means of the quantities needed for model atmosphere generation, which we then used to create the model atmosphere.\par

In the absence of on-site air pressure measurements, we devised a surface pressure estimation approach requiring only the site latitude, \Slat, longitude, \Slon~and altitude (height above sea level) \Salt~as inputs. The method employs the MERRA-2 vertical pressure levels, $lev$, and the mean edge heights, $H$, the mean altitudes to which they correspond, from the surrounding grid points. For each grid point, the closest mean edge heights above and below \Salt~are determined, the corresponding pressure levels identified and a linear interpolation between the pair of points is carried out to estimate the pressure at the height \Salt. The estimated pressures at \Salt~from the four grid points are then interpolated bilinearly to estimate the surface pressure, \Psurf, at the site location, \Sloc. Finally, the mean vertical profiles of other quantities (e.g. Figure 1 of \cite{Pesce2024}) are determined by bilinearly interpolating the mean MERRA-2 profiles of the surrounding grid points to (\Slat, \Slon), truncating the profiles at the closest vertical pressure level above \Salt~and using \Psurf~to logarithmically extrapolate them to \Sloc. To avoid negative wind speed predictions, wind speeds were instead estimated by linearly interpolating between the values at the pressure levels closest to \Psurf~above and below.\par

Following atmospheric model generation, we used the \am~atmospheric model \cite{Paine2019} to solve the radiative transfer equation for the models, effectively estimating ground-based radiometer measurements by simulating the effect of atmospheric absorption, emission, and scattering on radio-frequency signals passing through the model atmospheres. We chose a frequency range of $80$\,GHz to $700$\,GHz with $500$-MHz resolution. Estimated quantities include the mean precipitable water vapour (PWV), zenith optical depth $\tau(\nu)$, zenith brightness temperature $T_b(\nu)$, and wind speed $v_{\mathrm{wind}}$.

\section{Sensitivity Estimation}\label{sect:sensitivity_estimation}
Equipped with the $\tau(\nu)$ and $T_b(\nu)$ values, we estimated the effective sensitivity of the antenna and receiver at 230\,GHz through single-dish SEFDs for different dish diameters using the definition \cite{Taylor1999}
\begin{equation}
    \mathrm{SEFD} \equiv \frac{2k_\mathrm{B}T_{\mathrm{sys}}}{\eta A_{\mathrm{geo}}}. \nonumber
\end{equation}
In the equation above, $k_\mathrm{B}$ is the Boltzmann constant, $T_{\mathrm{sys}}$ the absorption-corrected system temperature, $A_{\mathrm{geo}}$ the geometric area of the dish and $\eta$ an efficiency encapsulating all relevant effects. Assuming a perfectly sideband-separating receiver system, we calculated $T_{\mathrm{sys}}$ as
\begin{equation}
    T_{\mathrm{sys}}(\nu) = e^{\tau(\nu)} [T_{\mathrm{rec}}(\nu) + (1 - e^{-\tau(\nu)})T_{\mathrm{atm}}(\nu)], \nonumber
\end{equation}
where $T_{\mathrm{rec}}(\nu)$ is the receiver temperature and $T_{\mathrm{atm}}(\nu)$, the effective atmospheric temperature, is
\begin{equation}
    T_{\mathrm{atm}}(\nu) = \frac{T_{b}(\nu) - T_{\mathrm{CMB}}e^{-\tau(\nu)}}{1 - e^{-\tau(\nu)}}. \nonumber
\end{equation}
$T_{b}(\nu)$ is the zenith atmospheric brightness temperature. We assumed the ALMA 230-GHz receiver temperature of $T_{\mathrm{rec}} = 40$\,K (see Table 5 of \cite{Pesce2024}) and $T_{\mathrm{CMB}}=2.725$\,K \cite{Mather1999}. Our $\eta$ estimates considered only the forward efficiency and aperture efficiency:
\begin{equation}
    \eta(\nu) = \eta_{\mathrm{F}}(\nu) \eta_{\mathrm{ap}}(\nu). \nonumber
\end{equation}
We chose $\eta_{\mathrm{F}} = 0.9$ for all sites, which is similar to values adopted in other EHT expansion analyses \cite{Pesce2024}. The aperture efficiency, $\eta_{\mathrm{ap}}$, was calculated using Ruze's formula \cite{Ruze1966}, assuming a focus offset of $10\,\mu$m and a commercially available dish with an RMS surface accuracy of $40\,\mu$m.\par

These calculations underpin the 230-GHz noise performance analysis of hypothetical dishes at the candidate sites in Section~\ref{sect:results_and_discussion}, incorporating dish and receiver specifications and the site-specific atmospheric conditions calculated in Section~\ref{sect:climatological_site_characterisation}. Three indicative dish diameters (9\,m, 13\,m and 18\,m) were included based on ngEHT and next-generation Very Large Array designs.

\section{Results and Discussion}\label{sect:results_and_discussion}
We validated the climatological characterisation results through comparison with BMAC weather data from the \ngehtsim~ package, which was the only South African candidate site whose weather data was included in \cite{Raymond2021} and \ngehtsim. The results were consistent, yielding mean relative differences of $\sim4$\,\%, $\sim9$\,\% and $\sim17$\,\% for the mean 230-GHz zenith optical depth, mean PWV and mean wind speed, respectively, which are lower than the RMS of each parameter. With the validation completed, the analysis of all three candidate \sa~sites was undertaken.\par
\begin{figure}[h]
    \centering
    \includegraphics[width=\columnwidth]{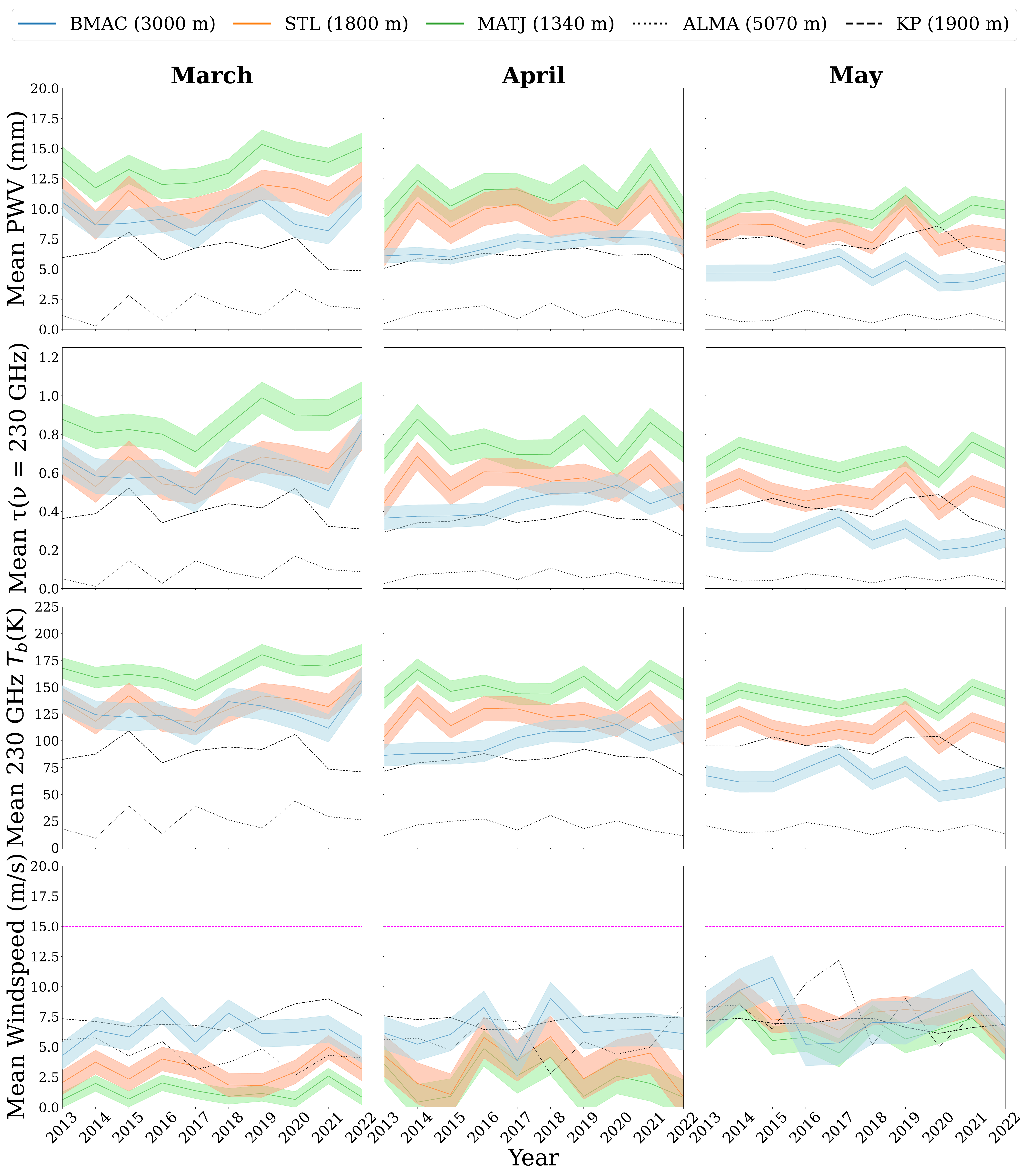}
    \caption{\small Monthly mean PWV (top row), 230-GHz zenith opacity (second row) and brightness temperature (third row), and wind speed (bottom row) above BMAC, STL, and MATJ for March, April, and May during 2013-2022. Predictions for BMAC, STL, and MATJ are shown by blue, orange, and green curves, respectively, with shaded regions indicating $\pm\sigma$. Means plus $\sigma$ above KP (dashed) and means minus $\sigma$ above ALMA (dotted) contextualise the \sa~values within the typical range of atmospheric conditions in the array. The dashed magenta line (bottom row) marks a wind speed of $15$\,m/s.}
    \label{fig:climatological_results_summary}
\end{figure}

\begin{figure}[h]
    \centering
    \includegraphics[width=\columnwidth]{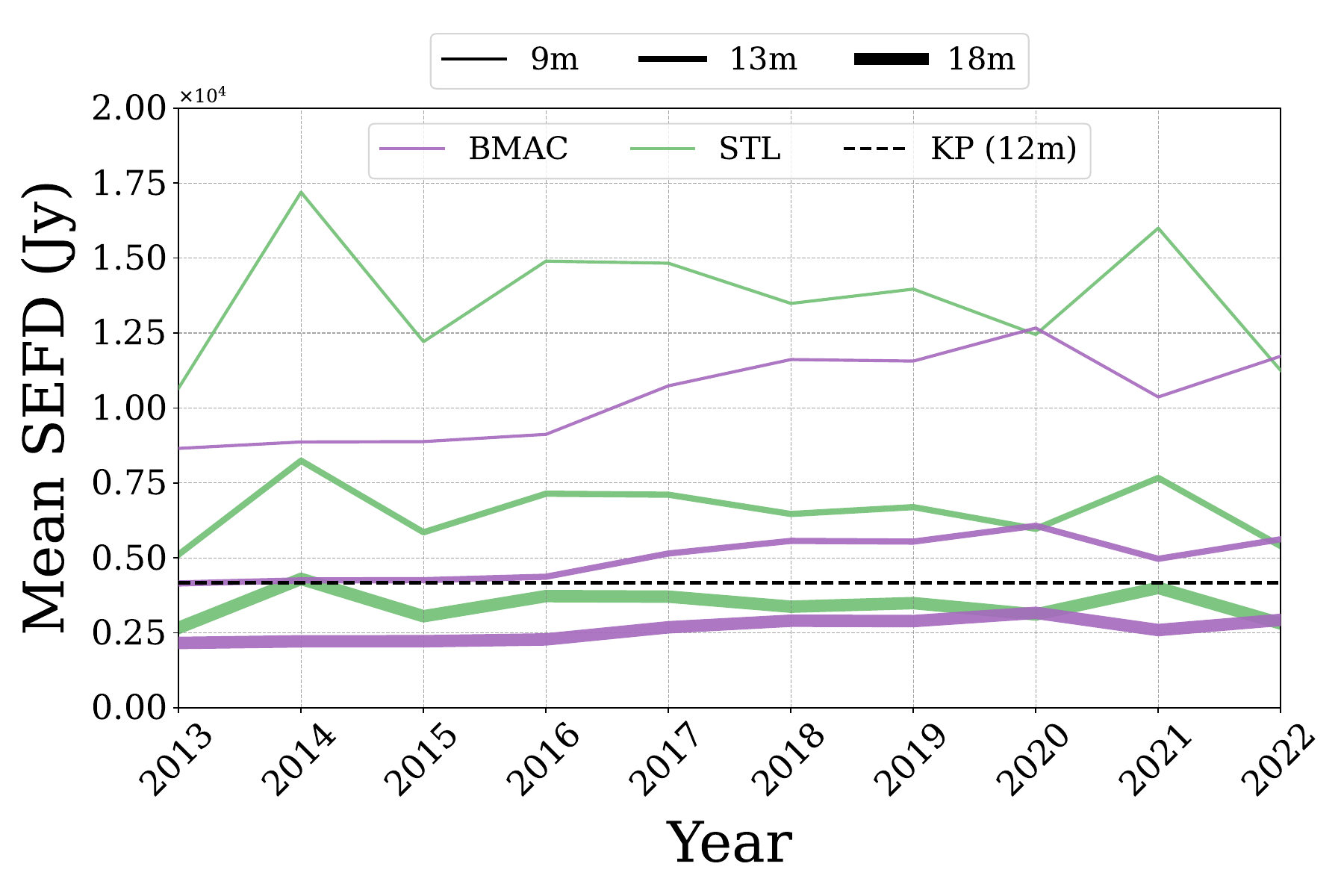}
    \caption{\small Mean SEFD estimates for different dish sizes during the month of April from 2013 to 2022. Thicker line widths correspond to large dish diameters at two South African sites, BMAC (purple) and STL (green). A black, dashed horizontal line at the mean value of KP over the decade is added for context. Values generally showed a standard deviation of $\sim$3000\,Jy.}
    \label{fig:sefd_comparison}
\end{figure}

\subsection{Nominal EHT observing window comparison}
Over the decade analysed, BMAC generally exhibited the lowest mean PWV values among the three South African sites. As an EHT site located at a relatively low altitude of 1900\,m, Kitt Peak (KP) has some of the least favourable mm observing conditions in the array. Thus, we used it as a benchmark non-premier site against which to evaluate the \sa~candidate sites. In April, the customary EHT observing window, the mean PWV and the 230-GHz optical depths and brightness temperatures above BMAC are comparable to those above KP (see middle column of Figure~\ref{fig:climatological_results_summary}), a pattern also observed at 86\,GHz and 345\,GHz. While STL's mean PWV, $\tau(\nu)$ and $T_b(\nu)$ curves approach those of BMAC during certain years, the STL and MATJ curves generally exceed the upper bound of the KP one-standard deviation ($\sigma$) range. In May, BMAC's curves fall well below $\mathrm{KP}+\sigma$, while those of STL remain close to them. The $\sim10$-mm conditions at BMAC and STL in March decline to $\sim5$\,mm (STL) and $<5$\,mm (BMAC) from April through August, representing an important trend as the ngEHT project seeks to extend the observing window to $\sim3$ months in Phase 1 \cite{Doeleman2023}.\par
\begin{figure}[h]
    \centering
    \includegraphics[width=\columnwidth]{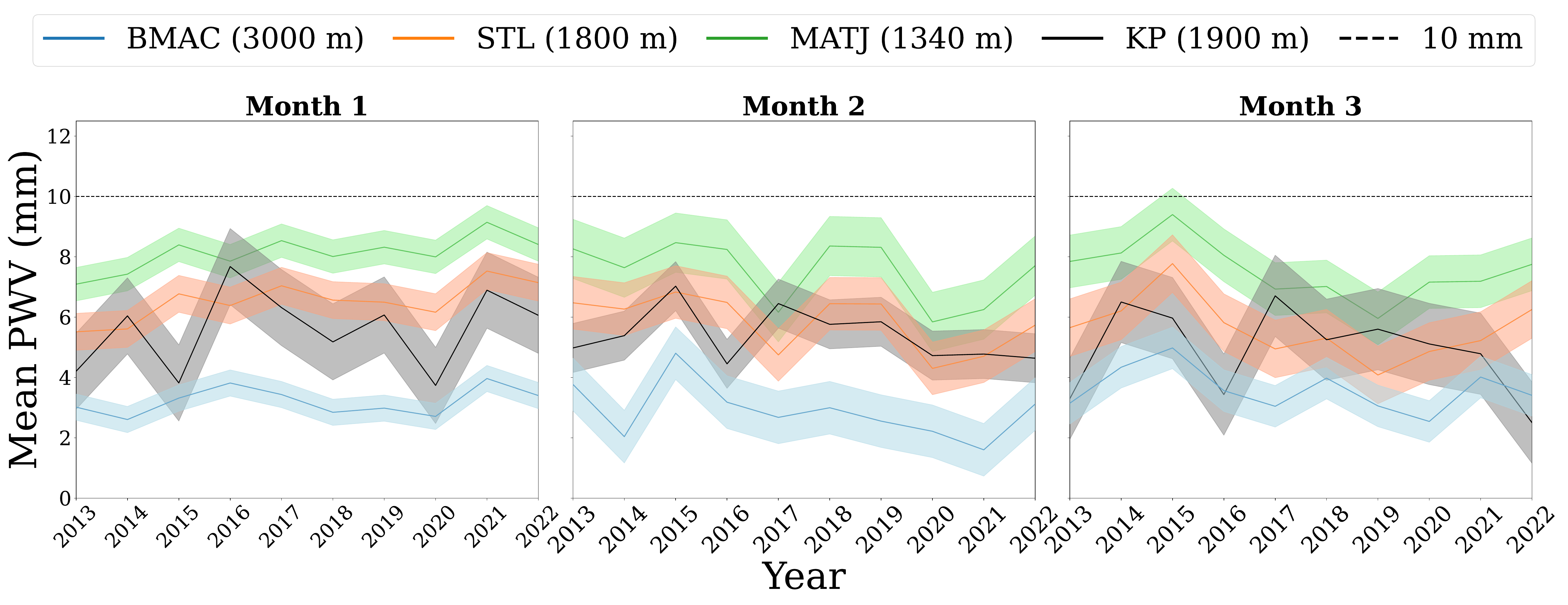}
    \caption{\small Winter monthly mean PWV above BMAC, STL, MATJ and KP during 2013-2022. Predictions for BMAC, STL, MATJ and KP are shown by blue, orange, green and black curves, respectively, with shaded regions indicating $\pm\sigma$. The dashed black line marks a PWV of 10\,mm.}
    \label{fig:winter-winter_comparison}
\end{figure}
Similar wind speeds to STL and MATJ during the first half of the year and significantly higher values during the latter half make BMAC the windiest site of the three. Wind-induced sensitivity degradation depends on antenna specifications but at wind speeds $<15$\,m/s, degradation is typically insubstantial \cite{Pesce2024}. The mean wind speeds at all three candidate \sa~sites rarely reach this mark. We note, however, that the MERRA-2 wind speed data lack the time-resolution to capture gusting. This is left for future analysis of on-site measurements.\par

We explored three dish diameters in our estimated sensitivity analysis: 9\,m, 13\,m and 18\,m. The hypothetical 13-m dish at BMAC and 18-m dish at STL showed mean SEFDs similar to those of the KP antenna in April (see Figure~\ref{fig:sefd_comparison}). Specifically, KP marginally outperforms a BMAC 13-m dish and is marginally outperformed by an STL 18-m dish, reflecting KP's slightly superior April 230-GHz zenith opacities and brightness temperatures. The noise performance difference between BMAC and STL diminishes with increasing dish diameter. While an 18-m dish at BMAC outperforms the same dish at STL, the difference is minimal during the latter half of the decade analysed, partly due to an apparent uptrend in April brightness temperatures and zenith opacities at BMAC. This trend warrants further monitoring with a MERRA-2 data subset including 2023 and 2024.

\subsection{Winter-winter comparison}
The goal of achieving year-round observations in Phase 2 of the ngEHT project \cite{Doeleman2023} relaxes the requirement for sub-10\,mm PWV conditions specifically around April, allowing a broader assessment of site quality across seasons. This enables us to remove the effect of opposing seasonal patterns between the southern and northern hemispheres by comparing the quality indicators of the candidate \sa~sites and KP during equivalent seasonal periods. Since winter observing conditions are generally more favourable than summer conditions, we focused on the meteorological winters: 1 June to 31 August in the southern hemisphere and 1 December to 28/29 February in the northern hemisphere. Figure~\ref{fig:winter-winter_comparison} demonstrates that STL and KP offer similar site quality for mm observations, while the higher-altitude BMAC site exhibits conditions superior to both. Typically falling below the 10\,mm mark in the same figure, all three \sa~candidate sites become competitive mm sites in the southern hemisphere winter. Notably, only in March and the period plotted in Figure~\ref{fig:winter-winter_comparison} do KP's mean PWVs and zenith spectral opacities consistently suggest more favourable conditions than BMAC and STL. Furthermore, observing conditions at the South African sites are relatively stable throughout the year. For instance, PWV values above KP deteriorate by a factor $>6$ (to $\sim30\,$mm in July and August) in the northern hemisphere summer compared to April, while even the MATJ PWV values seldom exceed $15$\,mm. This stability is significant for year-round observing and opens the door to southern-winter VLBI observations with the \amt~and Atacama Desert stations (ALMA and APEX), particularly with Frequency Phase Transfer (FPT)-enabled stations.

\section{Conclusion}\label{sect:conclusion}
Among the three candidate \sa~sites selected for desktop evaluation, BMAC showed the most promising proxies for mm site quality, comparable to KP during the nominal April EHT observing window and outperforming it during most of the year. In its most favourable years, however, STL's weather conditions are similar to those of BMAC in this window, and it is comparable to KP when the sites' weather conditions are compared for similar meteorological seasons. In our view, BMAC and STL justify the financial investment of in situ characterisation.\par

The positive mm site quality and antenna sensitivity indicators emphasise the importance of the next steps. FPT will very likely play a pivotal role in future EHT operations, particularly at non-premier sites. It is, therefore, also important to incorporate this technique in an extension of this study to dynamical imaging, especially characterising its performance as a function of sensitivity to help inform the recommended \sa~dish size.

\section{Acknowledgements}
We thank Lindy Blackburn, Iniyan Natarajan, Daniel Palumbo and Dominic Pesce for insightful discussions that greatly benefitted this work. We are especially grateful to Dominic Pesce for adding our candidate sites to \ngehtsim, which significantly streamlined the analysis. We also appreciate the constructive feedback from two anonymous referees.

%
%
\noindent\small

\begin{thebibliography}{9}
  
    \bibitem{Johnson2023}
    M. D. Johnson, K. Akiyama, L. Blackburn, K. L. Bouman, A. E. Broderick et al., 
    ``Key Science Goals for the Next-Generation Event Horizon Telescope,'' 
    \textit{Galaxies}, \textbf{11}, 3, April 2023, p. 61.

    \bibitem{Doeleman2023}
    S. S. Doeleman, J. Barrett, L. Blackburn, K. L. Bouman, A. E. Broderick et al., 
    ``Reference Array and Design Consideration for the Next-Generation Event Horizon Telescope,'' 
    \textit{Galaxies}, \textbf{11}, 5, October 2023, p. 107.

    \bibitem{Bower2015}
    G. C. Bower, J. Dexter, S. Markoff, M. A. Gurwell, R. Rao, and I. McHardy, 
    ``A Black Hole Mass-Variability Timescale Correlation at Submillimeter Wavelengths,'' 
    \textit{The Astrophysical Journal Letters}, \textbf{811}, 1, 2015, L6.

    \bibitem{Genzel2003}
    R. Genzel, R. Sch{\"o}del, T. Ott, A. Eckart, T. Alexander et al., 
    ``Near-infrared flares from accreting gas around the supermassive black hole at the Galactic Centre,'' 
    \textit{Nature}, \textbf{425}, 6961, 2003, pp. 934-937.
    
    \bibitem{Ghez2004}
    A. M. Ghez, S. A. Wright, K. Matthews, D. Thompson, D. Le Mignant et al., 
    ``Variable infrared emission from the supermassive black hole at the center of the Milky Way,'' 
    \textit{The Astrophysical Journal}, \textbf{601}, 2, 2004, L159.
    
    \bibitem{Gravity2020}
    Gravity Collaboration et al., 
    ``The flux distribution of Sgr A," 
    \textit{Astronomy \& Astrophysics}, \textbf{638}, 2020, A2.

    \bibitem{EHT2019_v} 
    Event Horizon Telescope Collaboration et al., 
    ``First M87 Event Horizon Telescope Results. V. Physical Origin of the Asymmetric Ring,'' 
    \textit{The Astrophysical Journal Letters}, \textbf{875}, 1, Apr. 2019, L5.

    \bibitem{EHT2022_i}
    Event Horizon Telescope Collaboration et al., 
    ``First Sagittarius A* Event Horizon Telescope Results. I. The Shadow of the Supermassive Black Hole in the Center of the Milky Way,'' 
    \textit{The Astrophysical Journal Letters}, \textbf{930}, 2, May 2022, L12.

    \bibitem{Deane2022}
    R. Deane and I. Natarajan, 
    ``The First Image of the Milky Way’s Central Black Hole and the Unique Enhancement Africa Could Offer Future Tests of Gravity,'' 
    \textit{South African Journal of Science}, \textbf{118}, 7-8, July 2022, pp. 1-4.
    
    \bibitem{LaBella2023}
    N. La Bella, S. Issaoun, F. Roelofs, C. Fromm, and H. Falcke, 
    ``Expanding Sgr A* Dynamical Imaging Capabilities with an African Extension to the Event Horizon Telescope,'' 
    \textit{Astronomy \& Astrophysics}, \textbf{672}, 2023, A16.

    \bibitem{Palumbo2023}
    D. C. M. Palumbo, G. N. Wong, A. Chael, and M. D. Johnson, 
    ``Demonstrating Photon Ring Existence with Single-baseline Polarimetry,'' 
    \textit{The Astrophysical Journal Letters}, \textbf{952}, 2, August 2023, L31.
    
    \bibitem{Cornwell1999}
    T. Cornwell and E. B. Fomalont,
    ``Self-Calibration,''
    in \textit{Synthesis Imaging in Radio Astronomy II}, eds. G. B. Taylor, C. L. Carilli, and R. A. Perley,
    \textit{Astronomical Society of the Pacific Conference Series}, \textbf{180}, January 1999, pp. 187-199.    

    \bibitem{Narayan1986}
    R. Narayan and R. Nityananda, 
    ``Maximum entropy image restoration in astronomy,'' 
    \textit{Annual Review of Astronomy and Astrophysics}, \textbf{24}, 1986, pp. 127-170.



    \bibitem{Roelofs2020}
    F. Roelofs, M. Janssen, I. Natarajan, R. Deane, J. Davelaar et al., 
    ``SYMBA: An End-to-End VLBI Synthetic Data Generation Pipeline - Simulating Event Horizon Telescope Observations of M 87,'' 
    \textit{Astronomy \& Astrophysics}, \textbf{636}, 2020, A5.

    \bibitem{Raymond2021}
    A. W. Raymond, D. Palumbo, S. N. Paine, L. Blackburn, R. Córdova Rosado et al., 
    ``Evaluation of New Submillimeter VLBI Sites for the Event Horizon Telescope,'' 
    \textit{The Astrophysical Journal Supplement Series}, \textbf{253}, 1, March 2021, p. 5.

    \bibitem{Yu2023}
    W. Yu, R. S. Lu, Z. Q. Shen, and J. Weintroub, 
    ``Evaluation of a Candidate Site in the Tibetan Plateau towards the Next Generation Event Horizon Telescope,'' 
    \textit{Galaxies}, \textbf{11}, 1, February 2023.

    \bibitem{Pesce2024}
    D. W. Pesce, L. Blackburn, R. Chaves, S. S. Doeleman, M. Freeman et al., 
    ``Atmospheric Limitations for High-Frequency Ground-Based VLBI,'' 
    \textit{arXiv e-prints}, April 2024, arXiv:2404.01482.

    \bibitem{Gelaro2017}
    R. Gelaro, W. McCarty, M. J. Suárez, R. Todling, A. Molod et al., 
    ``The Modern-Era Retrospective Analysis for Research and Applications, Version 2 (MERRA-2),'' 
    \textit{Journal of Climate}, \textbf{30}, 14, 2017, pp. 5419-5454.

    \bibitem{Paine2019}
    S. Paine, 
    ``The am Atmospheric Model,'' 
    Zenodo, March 2018, doi: 10.5281/zenodo.1193771.

    \bibitem{Taylor1999}
    G. B. Taylor, C. L. Carilli and R. A. Perley,
    ``Synthesis Imaging in Radio Astronomy II,''
    \textit{Astronomical Society of the Pacific Conference Series}, \textbf{180}, January 1999.

    \bibitem{Mather1999}
    J. C. Mather,  D. J. Fixsen, R. A. Shafer, C. Mosier and D. T. Wilkinson,
    ``Calibrator Design for the COBE$^*$ Far Infrared Absolute Spectrophotometer (FIRAS),''
    \textit{The Astrophysical Journal}, \textbf{512}, 2, February 1999, p. 511.

    \bibitem{Ruze1966}
    J. Ruze,
    ``Antenna Tolerance Theory -- A Review,''
    \textit{Proceedings of the IEEE}, \textbf{54}, 4, April 1966.

    



    
    
\end{thebibliography}
\end{document}